\documentclass[aps,pra,twocolumn,showpacs,superscriptaddress,shortbibliography]{revtex4-2}
\usepackage{graphicx} 
\usepackage{amsmath}
\usepackage{soul}
\usepackage{graphicx,epstopdf}
\usepackage{gensymb}
\usepackage[dvipsnames]{xcolor}
\usepackage{footnote}
\usepackage{multibib}

\newcommand{\be}{\begin{equation}}
	\newcommand{\ee}{\end{equation}}
\newcommand{\bea}{\begin{eqnarray}}
	\newcommand{\eea}{\end{eqnarray}}
\newcommand{\bse}{\begin{subequations}}
	\newcommand{\ese}{\end{subequations}}

\graphicspath{{../}}
\usepackage{color}
\usepackage[colorlinks,bookmarks=false,citecolor=darkblue,linkcolor=red,urlcolor=blue]{hyperref}

\definecolor{darkred}{rgb}{0.7,0.0,0.0}

\definecolor{darkblue}{rgb}{0,0.02,0.45}

\definecolor{darkgreen}{rgb}{0.02,0.45,0.0}

\definecolor{violet}{rgb}{0.8,0.2,0.6}

\begin{document}
\title{Magnetic ground state, critical analysis of magnetization, and large magnetocaloric effect in the ferromagnetically coupled kagome lattice YCa$_3$(MnO)$_3$(BO$_3$)$_4$}
\author{A. Choudhary}
\affiliation{School of Physics, Indian Institute of Science Education and Research Thiruvananthapuram-695551, India}
\author{A. Magar}
\affiliation{School of Physics, Indian Institute of Science Education and Research Thiruvananthapuram-695551, India}
\author{S. Ghosh}
\author{S. Kanungo}
\affiliation{School of Physical Sciences, Indian Institute of Technology Goa-403401, India}
\author{R. Nath}
\email{rnath@iisertvm.ac.in}
\affiliation{School of Physics, Indian Institute of Science Education and Research Thiruvananthapuram-695551, India}

\begin{abstract}
We report a detailed study of the magnetic properties, critical analysis of magnetization, and magnetocaloric effect of a spin-$2$ kagome lattice YCa$_3$(MnO)$_3$(BO$_3$)$_4$. The experiments are complemented by the density functional band structure calculations. The magnetic measurements suggest a highly frustrated nature of the compound due to competing ferro- and antiferromagnetic interactions with the dominant one being ferromagnetic. It undergoes a unconventional ferromagnetic ordering at $T^* \simeq 7.8$~K and a field induced metamagnetic transition in low fields, implying spin canting. A $H$-$T$ phase diagram is constructed that features three phase regimes. Indeed, the band structure calculations reveal dominant ferromagnetic interaction along the chains that are coupled antiferromagnetically yielding a frustrated kagome geometry. This compound shows a large magnetocaloric effect with isothermal entropy change $\Delta S_{\rm m} \simeq 12$~J/kg-K, adiabatic temperature change $\Delta T_{\rm ad} \simeq 8.4$~K, and relative cooling power $RCP \simeq 349$~J/kg for a field change of 7~T. The critical analysis of magnetization and magnetocaloric parameters suggests that the transition is a second order phase transition and it is tricritical mean field type. Owing to it's large magnetocaloric parameters, second order phase transition, and no thermal hysteresis, YCa$_3$(MnO)$_3$(BO$_3$)$_4$ emerges as a potential rare-earth free material for magnetic refrigeration.
\end{abstract}

\maketitle
\section{Introduction}
For decades, geometrically frustrated magnets in two-dimension have lured enormous attention, especially to explore the exotic phases of matter~\cite{lacroix2011}. A classic example of the 2D frustrated magnet is the kagome lattice, composed of corner-shared triangles with coordination number four~\cite{Norman041002}.
The kagome lattices with Heisenberg antiferromagnetic (AFM) interaction and low spin values set a perfect avenue to realize quantum spin liquid (QSL), a highly entangled quantum state with no magnetic long-range order (LRO). Indeed, it is experimentally observed in well celebrated $S=1/2$ kagome lattice compounds like herbertsmithite [ZnCu$_3$(OH)$_6$Cl$_2$]~\cite{Olariu087202,Norman041002}. 
Subsequently, the endeavor to realize spin-liquid and other fascinating phenomenon has also been made in kagome lattices with high spin ($S > 1/2$). For instance, the $S=5/2$ jarosite compounds, $A$Fe$_3$(OH)$_6$(SO$_4$)$_2$ feature a spin-glass transition for $A$ = (H/D)$_3$O and an antiferromagnetic ordering for $A$ = Na and K~\cite{Wills2161,Wills325,Inami752,Yildirim214446}. On the contrary, muon spin relaxation ($\mu$SR) study on $S=3/2$ jarosite compound KCr$_3$(OH)$_6$(SO$_4$)$_2$ reveals a clear signature of persistent spin fluctuations down to ~25~mK~\cite{Keren6451}. Oddity with other jarosites, $S=1$ jarosite compounds $M$V$_3$(OH)$_6$(SO$_4$)$_2$ ($M$ = Na, Rb, Ag, Tl, and NH$_4$) have strong ferromagnetic interactions~\cite{Papoutsakis2647}. In the $S=2$ kagome compound CsMn$_3$F$_6$(SeO$_3$)$_2$, the electron spin resonance and $\mu$SR experiments reveal a cooperative paramagnetic behavior down to 0.3~K, despite a large Curie-Weiss temperature of $\sim 49$~K~\cite{Lee094439}.

Generally, more attention is focused on the kagome AFMs because of the quest for a QSL state, but kagome ferromagnets (FM) also have equally interesting aspects to look for. The kagome FMs demonstrate anomalous Hall effect, Nernst effect, and chiral edge state in several intermetallic compounds~\cite{Chen144410,Asabaeabf1467,Howard4269}. The insulating kagome FM is also predicted to demonstrate topologically non-trivial bands~\cite{Mook134409}. The experimental realization of flat magnon band and magnon Hall effect is reported only in the insulating $S=1/2$ kagome ferromagnet Cu[1,3-bdc]~\cite{Chisnell214403,Chisnell147201}. Therefore, despite non-frustrated nature, kagome ferromagnets hold promise in elucidating many fundamental aspects of magnetism and corroborating the theoretical predictions.

Moreover, kagome ferromagnets are also of increasing demand as magnetocaloric effect (MCE) candidates~\cite{Magar054076}. MCE is a green-energy and cost-effective solution in refrigeration in the sub-kelvin range over conventional cryogenics like liquid helium and nitrogen. Magnetic cooling down to sub-kelvin temperatures can be achieved by adiabatic demagnetization of materials having a giant MCE~\cite{Pecharsky44}. For efficient cooling, the material should exhibit a large isothermal entropy change for a small change in magnetic field at low temperatures and negligible magnetic hysteresis~\cite{Phan325}. Therefore, ferromagnetic insulators with second-order phase transition can be excellent MCE materials for cooling to low temperatures. The mineral gaudefroyite Ca$_4$(MnO)$_3$(BO$_3$)$_3$CO$_3$ is a rare example of naturally occurring perfectly kagome lattice. It has dominant ferromagnetic (FM) interactions and undergoes a glassy transition below $T_{\rm f} \simeq 10$~K~\cite{Rukang5260}. Additionally, it is reported to be an excellent MCE candidate for hydrogen liquification~\cite{Rukang5260}. Subsequently, a derivative of gaudefroyite, YCa$_3$($M$O)$_3$(BO$_3$)$_4$, ($M= $~Cr, Mn, V) family has been synthesized. They crystallise in a hexagonal crystal structure with $P6_3/m$ space group. The Cr analogue ($S=3/2$) doesn't show any magnet LRO down to 2~K~\cite{Wang7535}. Similarly, the V analogue ($S=1$) also exhibits a broad dispersionless excitation without a magnetic LRO down to 50~mK~\cite{Silverstein044006,Miiller1315}. Interestingly, both compounds have very large Curie-Weiss temperature, characterizing them as highly frustrated magnets. On the other hand, the Mn analogue [YCa$_3$(MnO)$_3$(BO$_3$)$_4$] is reported to undergo a non-trivial FM ordering below 7.5~K~\cite{Li172403}. The neutron powder diffraction experiments suggested that it is composed of edge-sharing ferromagnetic chains of MnO$_6$ octahedra along the $c$-axis which are interconnected antiferromagnetically via BO$_3$ groups to form a kagome lattice in the $ab$-plane. Below 7.5 K, the magnetic moments of the chains order to give a normal $q=0$ structure.

\begin{figure}
 \includegraphics[width=\columnwidth]{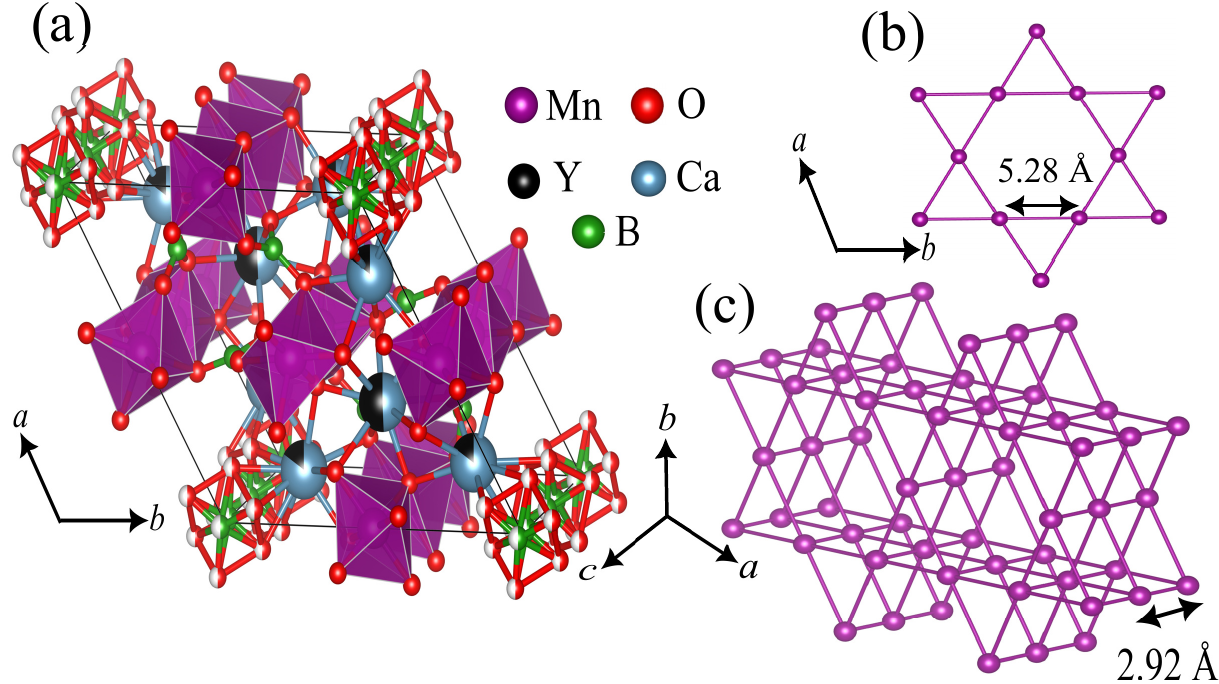}
\caption{\label{Fig1}(a) Unit cell of YCMBO featuring edge-shared MnO$_6$ octahedra. (b) A single kagome motif in the $ab$-plane made up of equilateral triangular units. (c) Two kagome planes separated by a relatively small distance of $2.92$~\AA, showing the formation of Mn$^{3+}$ chains along the $c$-axis.}
\end{figure} 
In this work, we present a detailed magnetic and magnetocaloric properties of a kagome ferromagnet YCa$_3$(MnO)$_3$(BO$_3$)$_4$ (YCMBO). In the crystal structure, as shown in Fig.~\ref{Fig1}, the MnO$_6$ octahedra are connected through BO$_3$ units to form a perfect kagome lattice in the $ab$-plane. The kagome layers are closely stacked via edge-sharing of MnO$_6$ octahedra forming continuous chains along the $c$-axis. While the interaction along the chain is found to be FM, the AFM interchain interaction constitute a kagome lattice. Along with a FM ordering at $T^* \simeq 7.8$~K, magnetization reveals a field induced meta-stable transition. It exhibits a large MCE comparable to other MCE materials having the transition temperature at around $T^*$. We also did a detailed critical analysis of magnetization and MCE parameters, establishing the second order nature and universality class of the phase transition.

\section{Methods}
A polycrystalline sample of YCMBO was prepared by the conventional solid-state reaction method. The starting materials used in this reaction were Mn$_2$O$_3$ (Sigma Aldrich, 99.9 \%), Y$_2$O$_3$ (Sigma Aldrich, 99.99\%), CaCO$_3$ (Sigma Aldrich, 99.9\%), and H$_3$BO$_3$ (Sigma Aldrich, 99.9\%). First, Y$_2$O$_3$ and CaCO$_3$ were pre-heated at 500$\degree$C for 6~hrs. Subsequently, the precursors taken in stoichiometric ratios were ground thoroughly and pressed into pellets. In order to compensate the loss of B in the high temperature synthesis, H$_3$BO$_3$ was taken 5\% extra. The pressed pellets were first fired at $900 \degree$C for $24$~hrs, followed by multiple firings at $1000\degree$C for $24$~hrs each with intermediate grindings. 

\begin{figure}
\includegraphics[width=\columnwidth]{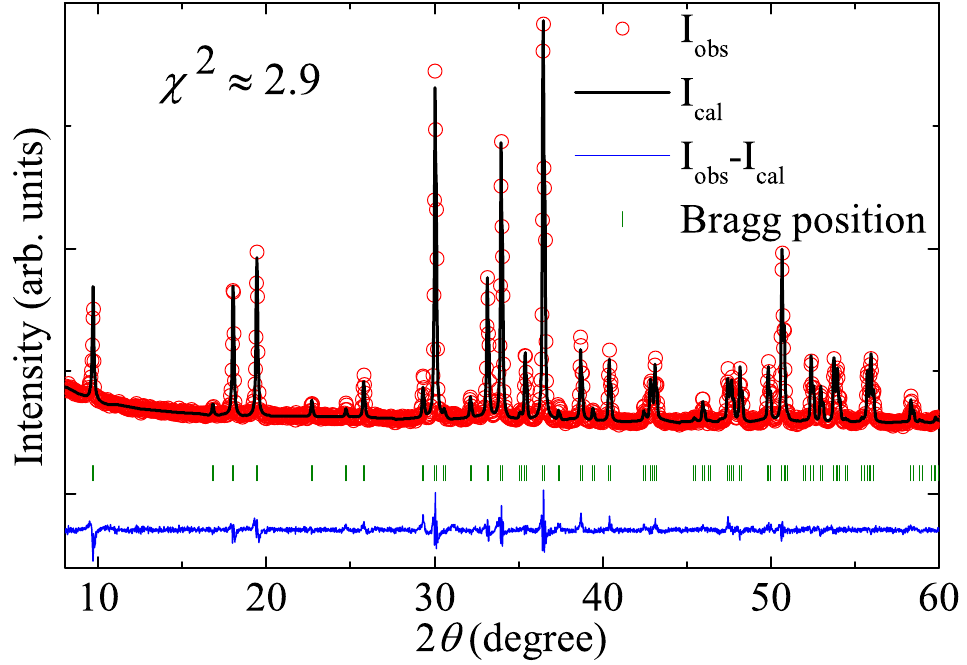}
\caption{\label{Fig2} Powder XRD data (circles) with the Rietveld fit (black solid line) of the titled compound. Bragg positions are shown as green vertical bars, and the bottom line indicates the difference between experimental and calculated intensities.}
\end{figure}

The phase purity of the polycrystalline sample was confirmed from powder x-ray diffraction (XRD) using PANalytical Xpert-Pro diffractometer (Cu~$K_\alpha$ radiation with $\lambda_{av}=1.54182$~\AA). The obtained powder XRD data is presented in Fig.~\ref{Fig2} along with the Rietveld fit. The Rietveld refinement was done using the FullProf software package, and initial structural parameters were adopted from the previous report~\cite{Li172403}. The refined lattice parameters $a = b= 10.553(8)$~\AA, $c = 5.847(7)$~\AA, $\alpha = \beta= 90^{\circ}$, and $\gamma = 120^{\circ}$ are in good agreement with the previous report. The absence of any extra peak and the reduced value of the goodness of fit ($\chi^2 \sim 2.9$) confirm the phase purity of the synthesized polycrystalline sample. The refined atomic co-ordinates are shown in Table~\ref{Table1}.

DC magnetization ($M$) as a function of temperature ($T$) and magnetic field ($H$) was measured using a superconducting quantum interference device (SQUID) (MPMS-3, Quantum Design) magnetometer in the $T$-range of 1.8~K to 380~K and in magnetic field up to 7~T.
AC susceptibility was measured as a function of temperature (2~K~$\leq T \leq 100$~K) and frequency (100~Hz~$ \leq \nu \leq 10$~kHz) in an AC field of $5$~Oe using the ACMS option of the PPMS. Heat capacity ($C_{\rm p}$) as a function of temperature (2~K~$\leq T \leq 300$~K) was measured on a small sintered pellet using the thermal relaxation technique in PPMS in different applied fields from 0 to 7~T.

The Electron Spin Resonance (ESR) experiments were performed using a Bruker EMX-series-X-band spectrometer ($v=9.4$~GHz). The sweeping of the magnetic field under a continuous microwave irradiation yields the ESR signal, which is recorded as the first field derivative of the absorbed microwave power $dP/dH$ as a function of the field.

The first-principles electronic structure calculations were carried out within the framework of density functional theory (DFT) employing the projector augmented-wave (PAW) method~\cite{Kresse558,Kresse11169} and the generalized gradient approximation (GGA) in the Perdew-Burke-Ernzerhof (PBE) formulation~\cite{Perdew3865} for the exchange-correlation functional, as implemented in the plane-wave-based Vienna Ab-initio Simulation Package (VASP)~\cite{Kresse558,Kresse11169}. A kinetic energy cutoff of 500~eV was used for the plane-wave basis set, which ensured convergence of the total energy. Structural optimization was performed by relaxing the internal atomic positions while maintaining the experimentally determined lattice parameters. The Mn atoms, occupying high-symmetry Wyckoff positions ($6g$), were constrained during relaxation using the selective dynamics scheme. The PBEsol~\cite{Perdew2544} functional was employed to improve the description of equilibrium structural properties, with convergence thresholds set to $10^{-5}$~eV for total energy and $10^{-3}$~eV/\AA~for the maximum Hellmann–Feynman force on each atom.

To incorporate on-site electronic correlations beyond standard GGA, the GGA+$U$ approach~\cite{Anisimov16929,Dudarev1505} was employed using an effective Hubbard parameter of $U_{\rm eff} =(U-J_H)= 5$~eV for the Mn-$d$ orbitals. The self-consistent calculations were performed with a stringent energy convergence criterion of $10^{-7}$~eV. The Brillouin zone (BZ) integration was performed using a $6 \times 6 \times 10$ Monkhorst–Pack $k$-point mesh. The effect of spin–orbit coupling (SOC) was included via relativistic corrections to the pristine Hamiltonian~\cite{Hobbs11556}; however, its influence on the electronic structure was found to be negligible for the present system. In addition to the electronic structure analysis, Bader charge analysis~\cite{Tang084204,Sanville899} was performed using the charge density from the converged self-consistent calculations to confirm the charge state of each atom.

\begin{table}
\caption{The Wyckoff positions, atomic coordinates, and occupancy of individual atoms in YCMBO, obtained from the Rietveld refinement of the powder XRD.}
\label{Table1}
\begin{ruledtabular}
\begin{tabular}{c c c c c c c }
 Atomic  & Wyckoff & $x$ & $y$ & $z$ & Occ\\
  sites&positions & & & & \\
  \hline
Mn&$6g$&0.5&0& 0&1&\\
Y(1)&$2c$&0.333(3)&0.666(7)& 0.25& 0.51&\\
Y(2)&$6h$&0.120(6)&0.837(9)& 0.25&0.16 &\\
Ca(1)&$2c$&0.333&0.666(1)& 0.25&0.49&\\
Ca(2)&$6h$&0.120(6)&0.837(9)& 0.25& 0.84 &\\
B(1)&$6h$&0.226(6)&0.789(6)& 0.75&1&\\
B(2)&$4e$&0&0&0.693&0.5&\\
O(1)&$6h$&0.086(5)&0.464(3)&0.25& 1&\\
O(2)&$6h$&0.315(1)&0.919(5)&0.75& 1&\\
O(3)&$12i$&0.299(1)&0.470(9)&0.547(6)& 1&\\
O(4)&$12i$&0.058(9)&0.914(2)&0.581(7)&0.5&\\
\end{tabular}
\end{ruledtabular}
\end{table}

\section{Results and Discussion}
\subsection{Magnetization}
\begin{figure*}
\includegraphics[width=\textwidth]{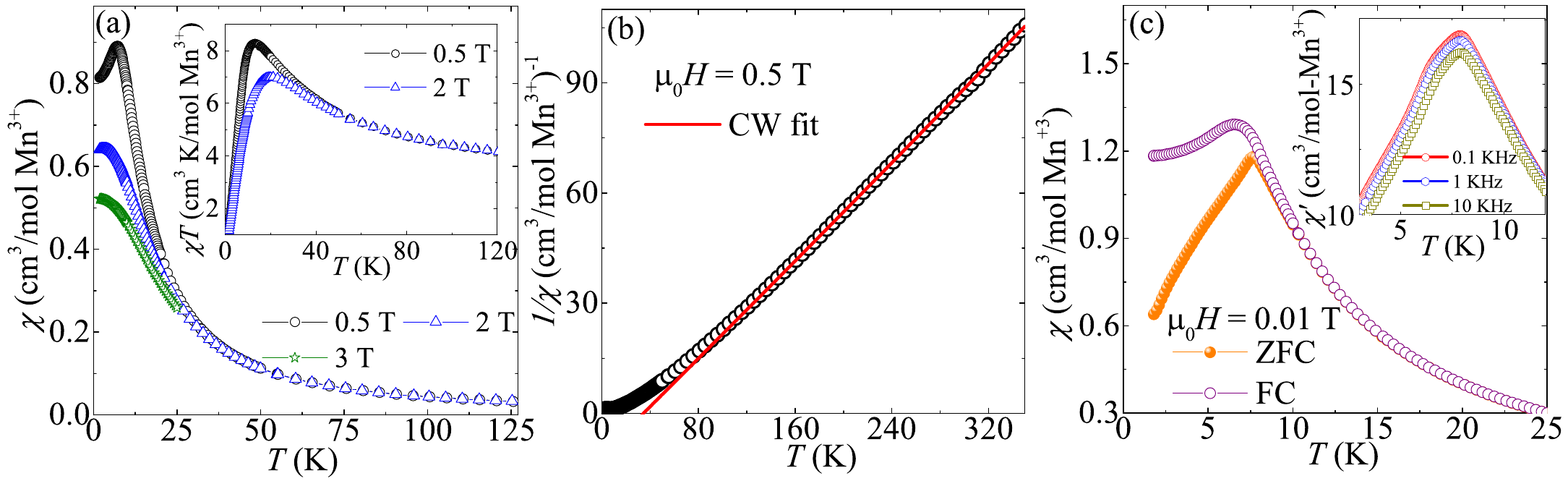}
\caption{\label{Fig3} (a) Magnetic susceptibility ($\chi$) as a function of temperature measured in different applied fields. Inset: $\chi T$ vs $T$ data. (b) Inverse susceptibility $1/\chi$ vs $T$ along with the CW fit (solid line). (c) $\chi (T)$ measured at $\mu_{0}H = 0.02$~T in ZFC and FC protocols. Inset: Real part of AC susceptibility ($\chi'$) measured in different frequencies.}
\end{figure*}

\begin{figure}
\includegraphics[width=0.9\columnwidth]{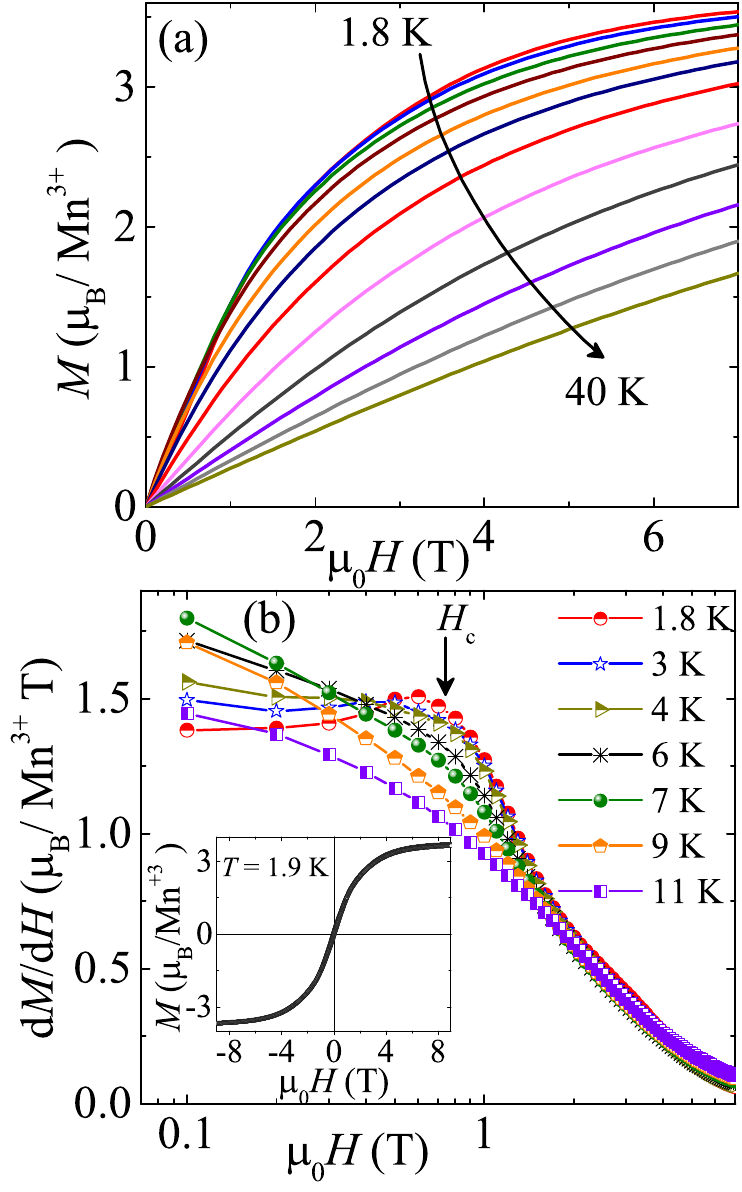}
\caption{\label{Fig4} (a) Isothermal magnetization ($M$) vs $H$ measured at different temperatures. (b) Temperature evaluation of the derivative of magnetization ($dM/dH$) exhibiting a slope change at low fields. Inset: Complete magnetization isotherm at $T=1.9$~K, showing saturation magnetization and minuscule hysteresis.}
\end{figure}
Temperature variation of magnetic susceptibility $\chi (\equiv M/H)$ measured in different applied magnetic fields is shown in Fig.~\ref{Fig3}(a). As the temperature decreases, $\chi$ increases systematically and shows a peak at $T^* \simeq 7.8$~K for $\mu_0 H = 0.5$~T, indicating a transition from paramagnetic to an ordered state. The peak-like feature shifts to lower temperatures with increasing field and becomes very broad above 2~T. To access the nature of interactions, we have plotted $\chi T$ vs $T$ in the inset of Fig.~\ref{Fig3}(a). $\chi T$ first increases, passes through a broad maximum, and then decreases towards low temperatures. This behavior points towards the presence of coexisting FM and AFM interactions in YCMBO~\cite{Savina104447,Mohanty134401}. This is indeed consistent with the findings from neutron diffraction experiments reported earlier~\cite{Li172403}.

To extract the magnetic parameters, $1/\chi$ vs $T$ data in the high temperature regime ($T \ge 150$~K)  were fitted by the Curie-Weiss (CW) law
\begin{equation}
\chi(T) = \chi_{0}+\frac{C}{(T-\theta_{\rm CW})},
\end{equation}
where $\chi_{0}$ is the $T$-independent susceptibility, $C$ is the Curie constant, and $\theta_{\rm CW}$ is the characteristic Curie-Weiss temperature. The fit shown in Fig.~\ref{Fig3}(b) gives $\chi_0 \simeq -1.6 \times 10^{-4}$~cm$^3$/mol-Mn$^{3+}$, $C\simeq  3.04$~cm$^3$K/mol-Mn$^{3+}$, and $\theta_{\rm CW} \simeq 34$~K. The fitting parameter $C$ gives the effective magnetic moment of $\mu_{\rm eff} \simeq 4.93~\mu_{\rm B}$ which is close to the spin only effective moment of Mn$^{3+}$ ($S=2$) ion [$(\mu_{\rm eff})_{\rm theory} = 4.89~\mu_{\rm B}$], assuming $g = 2$. Indeed, our ESR experiment (discussed later) yield $g\sim 2$. According to mean-field theory, $\theta_{\rm CW}$ is the sum of all exchange couplings in the system. Thus, the large and positive value of $\theta_{\rm CW}$ for YCMBO indicates that the dominant interaction is FM in nature.
Further, $T^* \sim 7.8$~K is considerably suppressed compared to $\theta_{\rm CW}$, yielding a frustration ratio of $f = \frac{|\theta_{\rm CW}|}{T^{*}} \simeq 4$. Such a value of $f$ is usually considered as moderate frustration in an AFM system. However, in YCMBO due to competing FM and AFM interactions, the value of $\theta_{\rm CW}$ is expected to be reduced, as compared to a purely AFM system~\cite{Nath024418}. Hence, $f$ does not reflect the actual strength of frustration in such compounds.

$\chi$ vs $T$ measured in a small magnetic field of 100~Oe in zero-field-cooled (ZFC) and field-cooled (FC) protocols exhibits a bifurcation below $T^*$ [see Fig.~\ref{Fig3}(c)]. To scrutinize the possibility of spin-glass below the ordering temperature, we have performed the frequency-dependent AC susceptibility measurements in a small AC field of 5~Oe around $T^*$, as shown in the inset of Fig.~\ref{Fig3}(c). The real part of AC susceptibility ($\chi'$) shows a peak at $T^*$, but it is completely frequency independent, suggesting the absence of spin freezing or spin-glass behavior. The bifurcation observed in ZFC and FC $\chi(T)$ might be due to a small amount of disorder in the polycrystalline sample~\cite{Dzara2810}.

Figure~\ref{Fig4}(a) presents isothermal magnetization curves from 1.8~K to 40~K in small temperature steps. A complete $M$ vs $H$ curve at $T = 1.9$~K is shown in the inset of Fig.~\ref{Fig4}(b). For $T=1.8$~K, magnetization increases slowly with an increase in field and almost saturates to a value, $M_{\rm s} \simeq 3.6$~$\mu_{\rm B}$ at 7~T, which is close to the expected saturation magnetization for a $S=2$ system ($M_{\rm s}=gS\mu_{\rm B}=4\mu_{\rm B}$). Although the magnetization reaches saturation, it is achieved in a relatively large field in contrast to a conventional FM. This suggests that a higher field was required to saturate the AFM interaction. Additionally, there is a slope change at around $\mu_0H\simeq 0.7$~T, as clearly visible in the $dM/dH$ vs $H$ plot shown in Fig.~\ref{Fig4}(b). Such a slope change at an intermediate field ($H_{\rm c}$) is not envisaged, indicating spin canting due to an AFM interaction competing with dominant FM interaction~\cite{Mohanty134401}. Further, the $M$ vs $H$ curve recorded at $T = 1.9$~K doesn't show any hysteresis [inset of Fig.~\ref{Fig4}(b)], supporting the absence of a glassy component. 

\subsection{Electron Spin Resonance}
\begin{figure}
\includegraphics[width=0.9\columnwidth]{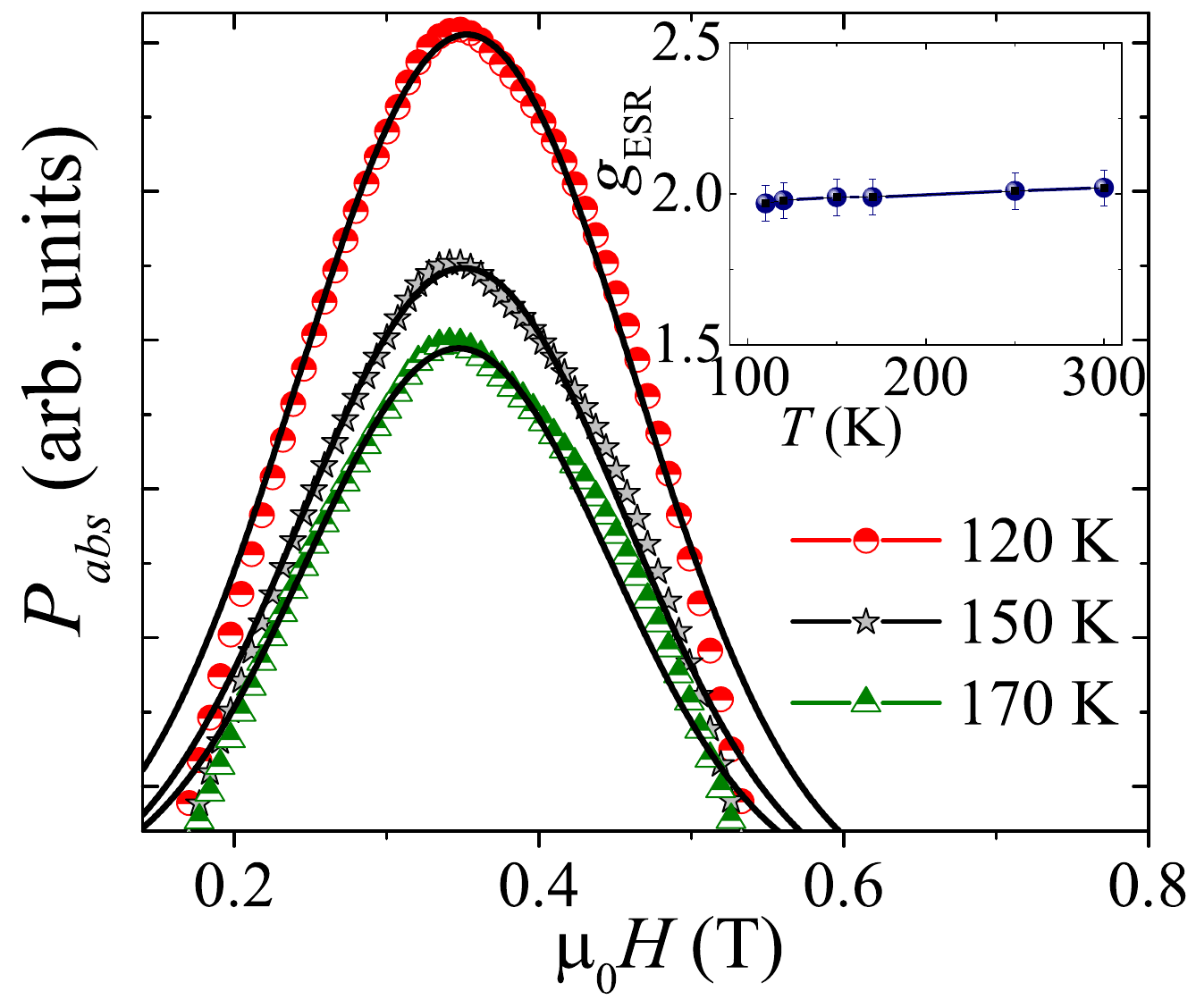}
\caption{\label{Fig5} Absorbed microwave power ($P_{\rm abs}$) vs $H$ measured at different temperatures. Solid lines are the fits using a Gaussian function. Inset: Variations of $g$-factor with temperature.}
\end{figure}
ESR spectra were recorded down to 110~K on the powder sample. Figure~\ref{Fig5} presents the actual absorbed microwave power ($P_{\rm abs}$) as a function of field at three different temperatures. $P_{\rm abs}$ shows a maxima at resonance field ($H_{\rm res}$). The exact value of $H_{\rm res}$ for each temperature is determined from a Gaussian fit to the spectra (solid line in Fig.~\ref{Fig5}). The Lande $g$-factor is then calculated using the resonance condition, $hv=g\mu_{\rm B}H_{\rm res}$ where, $h$ is the Planck's constant and $v$ is the resonance frequency corresponding to $H_{\rm res}$. We found $g \simeq 2$ in the entire measured temperature range as shown in the inset of Fig.~\ref{Fig5}.

\subsection{Heat Capacity}
\begin{figure}
\includegraphics[width=0.9\columnwidth]{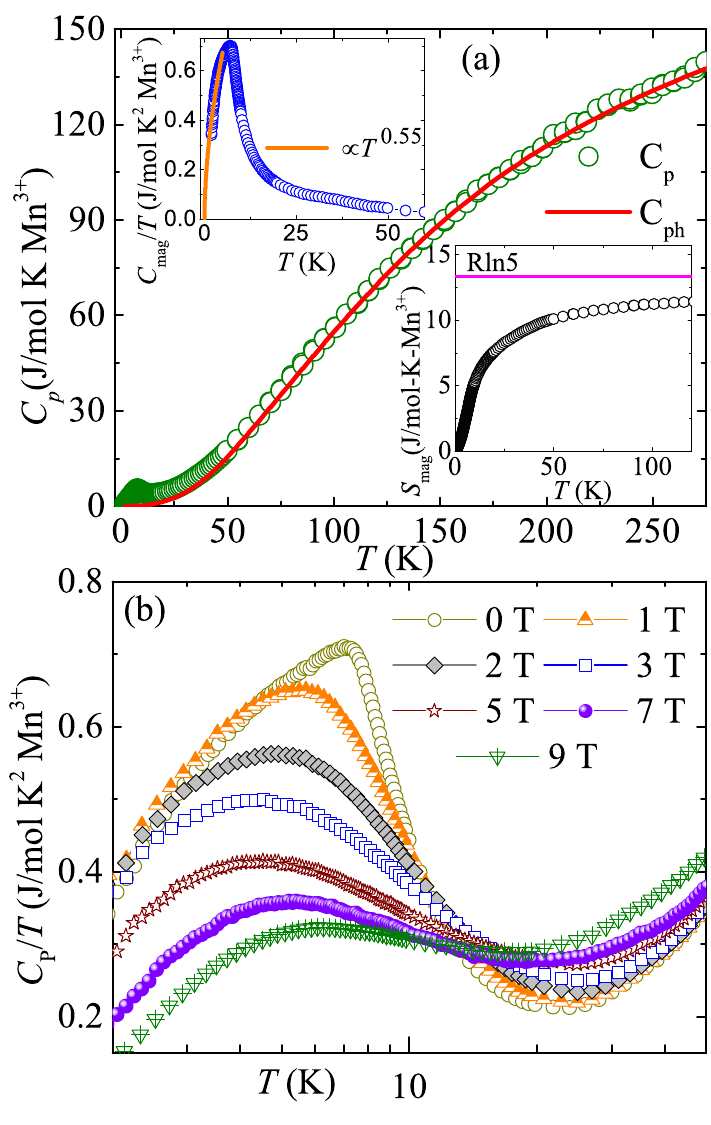}
\caption{\label{Fig6} (a) Heat capacity ($C_{\rm p}$) vs $T$ measured in zero-field. The solid line represents the phonon contribution ($C_{\rm ph}$) determined using the Debye-Einstein model.
Upper and lower insets show the magnetic heat capacity ($C_{\rm mag}/T$) and magnetic entropy ($S_{\rm mag}$) as a function of temperature in zero-field, respectively. (b) $C_{\rm p}/T$ vs $T$ in different magnetic fields in the low-temperature regime.}
\end{figure}
Temperature dependent heat capacity ($C_{\rm p}$) is presented in Fig.~\ref{Fig6}. The zero-field $C_{\rm p}$ data exhibit an anomaly at $T^{*} \simeq 7.5$~K, demonstrating the transition to a magnetic ordered state. YCMBO being a magnetic insulator, the major contributions to $C_{\rm p}$ are from the magnetic ($C_{\rm mag}$) and phonon ($C_{\rm ph}$) parts. The high temperature contribution is mostly $C_{\rm ph}$ while the low temperature contribution is dominant $C_{\rm mag}$. In order to estimate $C_{\rm mag}$, one needs to subtract $C_{\rm ph}$ from the total heat capacity. We approximated the phonon contribution by simulating the data by a linear combination of one Debye and four Einstein terms as~\cite{Kolay224405}
\begin{equation}
C_{\rm ph}(T)=f_{\rm D}C_{\rm D}(\theta_{\rm D},T)+\sum_{i = 1}^{4}g_{i}C_{{\rm E}_i}(\theta_{{\rm E}_i},T).
\label{Eq5}
\end{equation}
The first term in Eq.~\eqref{Eq5} is the Debye contribution to $C_{\rm ph}$, which can be written as 
\begin{equation}
C_{\rm D} (\theta_{\rm D}, T)=9nR\left(\frac{T}{\theta_{\rm D}}\right)^{3} \int_0^{\frac{\theta_{\rm D}}{T}}\frac{x^4e^x}{(e^x-1)^2} dx.
\label{Eq6}
\end{equation}
Here, $R$ is the universal gas constant, $\theta_{\rm D}$ is the characteristic Debye temperature, and $n$ is the number of atoms in the formula unit. The second term in Eq.~\eqref{Eq5} gives the Einstein contribution to $C_{\rm ph}$ that has the form
\begin{equation}
C_{\rm E}(\theta_{\rm E}, T) = 3nR\left(\frac{\theta_{\rm E}}{T}\right)^2 
\frac{e^{\theta_{\rm E}/T}}{[e^{\theta_{\rm E}/T}-1]^{2}}.
\label{Eq7} 
\end{equation}
Here, $\theta_{\rm E}$ is the characteristic Einstein temperature. The coefficients $f_{\rm D}$, $g_1$, $g_2$, and $g_3$ represent the fraction of atoms that contribute to their respective parts. These values are taken in such a way that their sum should be equal to one. The simulated phonon contribution is shown as a solid line in the main panel of Fig.~\ref{Fig6}(a) and the obtained $C_{\rm mag}/T$ is plotted as a function of temperature in the upper inset of Fig.~\ref{Fig6}(a).

The magnetic entropy is calculated as, $S_{\rm{mag}}(T) = \int_{\rm0\,K}^{T}\frac{C_{\rm {mag}}(T')}{T'}dT' \simeq 11.5$~J/mol-K [see the lower inset of Fig.~\ref{Fig6}(a)]. It is slightly lower than the expected magnetic entropy for a $S=2$ system, $S_{\rm mag}=R\ln 5=13.38$~J/mol-K. It is to be noted that only $\sim 40$\% of the total entropy is released just above $T^*$ and the remaining entropy is released at very high temperatures. This is unlike a conventional FM ordering where the entire entropy is expected to be released just above the transition, suggesting that YCMBO is not a simple FM and has a significant frustration effect. A power law, $C_{\rm p} = AT^n$ is fitted to $C_{\rm mag}/T$ below 5~K ($< T^*$) that yields $n \simeq 0.5$. Although the value of the exponent is close to the one expected for FM spin wave excitations~\cite{Gopal2012}, the ordering is far from a simple FM as understood from magnetic entropy and field dependent $C_{\rm p}$ discussed below.

Figure~\ref{Fig6}(b) presents $C_{\rm p}/T$ measured in different applied fields. The zero-field peak initially shifts toward low temperatures, similar to an AFM. With further increase in field ($\geq 5$~T), the peak becomes very broad and starts shifting to the high temperatures, as expected for a FM. 
This type of field evaluation also supports considerable spin canting in the compound.

\subsection{Phase Diagram}
\begin{figure}
\includegraphics[width=\columnwidth]{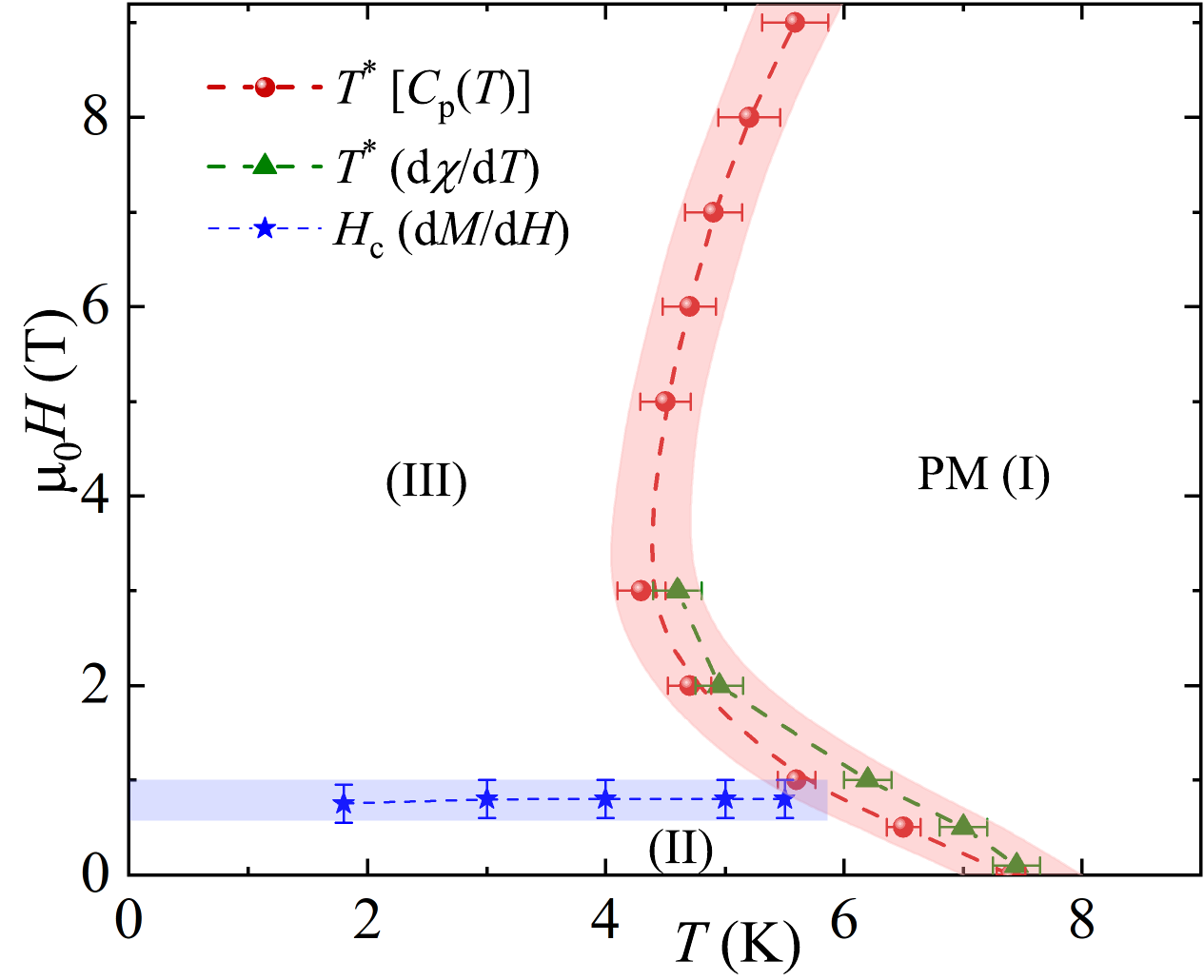}
\caption{\label{Fig7} $H$-$T$ phase diagram for YCMBO constructed from the magnetization and heat capacity measurements.}
\end{figure}
A $H$-$T$ phase diagram of YCMBO is made by plotting the field dependence of $T^*$ obtained from $\chi(T)$ and $C_{\rm p}(T)$ measurements along with $H_{\rm c}$ obtained from the magnetic isotherms. As depicted in Fig.~\ref{Fig7}, YCMBO undergoes a transition from PM (phase I) to an ordered state (phase II) in zero-field at $T^*$. As the field increases, $T^*$ moves towards low temperatures in small fields (up to $\sim 2$~T) and then shifts towards high temperatures in larger fields. Such a behaviour is very unusual and not commonly seen in case of pure ferromagnets. This can possibly be explained in terms of competing FM and AFM interactions. Typically, for a pure FM order, as the field increases, one would expect $T^*$ to shift to high temperatures only. However, in the present case, first, the AFM interaction get suppressed and saturates at about 2~T, leading to a shift of $T^*$ to the low temperature regime. For $H > 2$~T, only the ferromagnetic interaction is influenced, resulting in a outward movement of $T^*$ in higher fields. Further, a field-induced transition $H_{\rm c}$ is observed in isothermal magnetization that remains temperature independent. Similar field induced phases are typically reported for anisotropic magnets~\cite{Sebastian104425}. When plotted together, the phase diagram features three distinct phase regimes (I, II, and III). Though, such a phase diagram supports the presence of competing FM and AFM interactions, the exact nature of the phases would require further experiments including neutron diffraction in magnetic fields.


\subsection{Electronic Structure Calculations}
\begin{figure*}
    \centering
    \includegraphics[width=\textwidth]{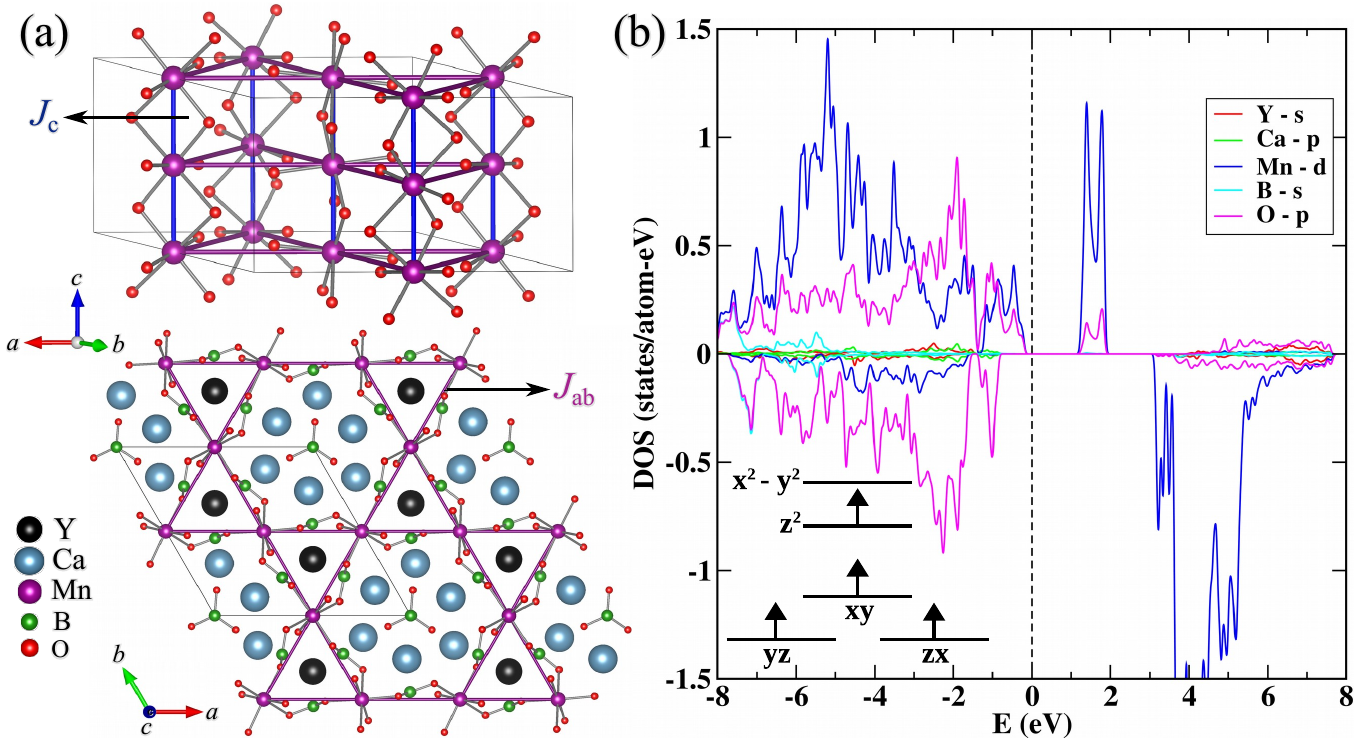}
    \caption{(a) Three-dimensional crystal structure of YCMBO is shown. Upper and Lower panels show the projection along the crystallographic $c$-axis and in the $ab$-plane, respectively, of the bulk structure. Two distinct magnetic exchange interactions between Mn atoms are highlighted ($J_{\rm c}$ and $J_{\rm {ab}}$). (b) Calculated spin-polarized GGA+$U$ DOS, projected onto atomic orbitals: Y-$s$, Ca-$p$, Mn-$d$, B-$s$, and O-$p$ states. The energy axis is referenced to the Fermi level ($E_\mathrm{F} = 0$~eV). Lower inset: schematic representation of the crystal-field splitting and occupancy of the Mn-$d$ orbitals.
    \label{Fig8}}
\end{figure*}

As mentioned earlier, YCMBO crystallizes in a hexagonal lattice, adopting the well-known gaudefroyite-type structure~\cite{Rukang5260}, and belongs to the space group $P6_3/m$ (No.~176). Experimental reports indicate a degree of site disorder involving Y, Ca, B, and O atoms; however, for the sake of calculation limitations, a disorder-free configuration with space group $P\bar{6}$ (No. 174) was considered, preserving the chemical stoichiometry intact as that of the experimentally synthesized compound. The crystal structure comprises infinite one-dimensional chains of edge-sharing MnO$_6$ octahedra extending along the $c$-axis. The Mn atoms occupy octahedral coordination and exhibit a characteristic Jahn–Teller distortion, with two elongated Mn–O bonds ($\sim$2.19~\AA) and shorter equatorial bonds ($\sim$1.90 and 1.99~\AA). These MnO$_6$ edge shared chains are bridged by planar triangular BO$_3$ units, giving rise to a kagome-like network within the $ab$-plane [see Fig.~\ref{Fig8}(a)(lower)], as a result, the Mn sublattice forms a kagome lattice in the $ab$-plane. This kagome geometry introduces geometric frustration, which is central to the material’s complex magnetic behavior. The three-dimensional framework formed by MnO$_6$ octahedra and BO$_3$ triangles contains tunnel-like interstitial channels running along the $c$-axis. These voids are occupied by Y and Ca cations, supported by the B and O atoms, contributing towards the stability of the crystal structure.

The spin-polarized density of states (DOS), calculated within the GGA+$U$ formalism and projected onto Y-$s$, Ca-$p$, Mn-$d$, B-$s$, and O-$p$ orbitals, is presented in Fig.~\ref{Fig8}(b). The DOS is plotted over a wide energy window ranging from $-8.0$~eV below the Fermi level to $8$~eV above the Fermi level, relative to the Fermi energy ($E_{\rm F}$). Close to the $E_{\rm F}$, the Mn-$d$ states dominate, where the majority spin channel states of Mn-$d$ are mostly filled, except a single unoccupied $e_g$ state, while the minority spin channel is completely empty. The calculated magnetic moment at the Mn site is  $3.91~\mu_B$ with total magnetic moment $12~\mu_B$/formula unit. The calculated DOS and magnetic moment together infer that Mn is in +3 oxidation state in the 3$d^4$ configuration with high spin $S=2$ state, as shown in the inset of Fig.~\ref{Fig8}(b). The Y and Ca are in nominal +3 and +2 oxidation states, with filled shell configurations, respectively. The B and O atoms are calculated to be in the +3 and -2 oxidation states, respectively, consistent with the charge neutrality. Interestingly, we found strong hybridization between O-2$p$ state with the Mn-3$d$ as reflected in the strong overlap of DOS. As a result substantial induced magnetic moment at the oxygen sites. These $d$-$p$ hybrid states dominate the vicinity of the Fermi level and contribute to the emergence of a band gap of approximately 1.2~eV in the majority spin channel, indicating an insulating ground state, while the contributions from Y, Ca, and B atoms are minimal near the $E_{\rm F}$.

As the Mn-sublattice forms an interesting kagome geometry in the $ab$ plane and chains along the $c$-axis, it is likely to have competition between the quasi-1D vs quasi-2D nature of the magnetic model. We calculated the magnetic ground state via comparative energetics of the FM and AFM spin configurations, and found that FM is energetically lower, thus establishing the FM ground state. Our calculated results are broadly consistent 
with the earlier report of the observed ferromagnetism in neutron diffraction measurements~\cite{Li172403}. To explore further details of the underlying spin model, we went ahead with the calculation of the magnetic exchange interactions between different Mn sites. For the sake of simplicity and to extract the essence of the spin model, we consider two exchange paths, intra-plane coupling ($J_{\rm ab}$) and out-of-plane i.e, intra-chain coupling ($J_{\rm c}$), as shown in Fig.~\ref{Fig8}(a). $J_{\rm c}$ is along the MnO$_6$ chains with Mn-Mn distance $\sim 2.93$~\AA~while $J_{\rm ab}$ is across the kagome layers where Mn-Mn distance is $\sim5.28$~\AA.

There are various methods to calculate the magnetic exchange interaction at the $ab$-$initio$ level. We have adopted the total energy method~\cite{Helberg3489,Mazurenko014418,Xiang224429}, where the DFT total energies of the different spin configurations are mapped to the corresponding Ising model with two exchange interactions, as shown in the model Hamiltonian 
\begin{equation}
H = J_c \sum_{\langle i,c \rangle} \mathbf{S^z}_i \mathbf{S^z}_{i+c} + J_{ab} \sum_{\langle i,ab \rangle} \mathbf{S^z}_i \mathbf{S^z}_{i+ab}.
\label{Hamil} 
\end{equation}
Here, $\mathbf{S^z}$ is the $z^{th}$ component of the spin operator, $\langle i,c \rangle$ and $\langle i,ab \rangle$ denote summation over nearest-neighbor spins along the chain and in the $ab$-plane, respectively. Although this method has a dependency on the exchange-correlation functional, choice of $U$, however, this method is successful in providing a qualitative estimate and understanding of the relative strength and sign of the exchange coupling~\cite{Sannigrahi245802,Roy085107,Roy155110,Kanungo161116}

The total energy calculations show that the unique structural motif of YCMBO leads to quasi-1D magnetic behavior. The edge shared MnO$_6$ chains along the $c$-axis exhibit strong FM exchange coupling with an intra-chain exchange constant of $\sim 2.3$~meV ($J_c/k_{\rm B}\simeq26.7$~K), while the intra-plane exchange interaction is AFM and relatively weak ($\sim0.25$~meV, which is $J_{ab}/k_{\rm B} \simeq2.9$~K). Such a strong FM coupling is rare for Mn–O–Mn interactions, which in many oxides like MnO or LiMnBO$_3$~\cite{Li052405} is typically AFM in nature in the iso-electronic and iso-spin configurations in a Jahn–Teller distorted environment. According to the Goodenough-Kanamori rule, the Mn-O-Mn super-exchange via 90\textdegree~favors FM interaction for the $d^4$ high spin $S=2$ systems~\cite{Goodenough1963,*Kanamori87}. A careful investigation of bond angles divulges that due to the distorted MnO$_6$ octahedra, the two $\angle$Mn-O-Mn angles in YCMBO are $\sim 83.5$\textdegree~and $\sim102.4$\textdegree~along the chain~\cite{Li052405}. 
Notably, the oxygen atoms bridging Mn atoms along the chain exhibit substantial magnetic moments, with values of –0.065/0.083 $\mu_B$ and 0.055/0.059 $\mu_B$ for short and long Mn–O–Mn bonds, respectively, in contrast to the negligible moment observed on oxygen atoms that connect adjacent chains. This spin polarization of the O atoms is a clear indication of strong Mn–O–Mn FM super-exchange interactions along the chains. Each linking oxygen atom, therefore, forms two bonds to adjacent Mn ions via either two half-occupied Mn $dz^2$ orbitals or two empty Mn $dx^2-y^2$ orbitals [Fig.~\ref{Fig8}(b)(lower)]. Furthermore, the relatively short Mn–Mn distances within the chains, compared to those in the $ab$-plane, support the conclusion that intra-chain super-exchange is the dominant magnetic interaction pathway.

In contrast, the inter-chain interactions within the kagomé plane are geometrically frustrated. Due to the triangular coordination of magnetic ions, a fully compensated AFM arrangement cannot be simultaneously satisfied among all nearest neighbors, leading to a large ground-state degeneracy. The extremely small value of the intra-plane interactions suggests that the coupling arises from higher-order mechanisms such as super-super-exchange via B and O or dipolar interactions, both of which are suppressed by the in-plane frustration. Consequently, magnetic LRO only sets in below a relatively low temperature of approximately 7.8~K, adopting a $q = 0$ propagation vector consistent with the calculated results~\cite{Li172403}.

The magnetocrystalline anisotropy energy (MAE) was evaluated as the total energy difference between the spin quantization oriented along the [001] (out-of-plane) and [110] (in-plane) crystallographic directions, defined as $E_{001} - E_{110}$. The resulting MAE is $\sim 33.18$ meV, indicating that the [001] direction is energetically favored over the [110] direction. The negative value of the anisotropy energy thus implies a uniaxial anisotropy with the easy magnetization axis aligned along the out-of-plane [001] direction.

A comparative analysis with isostructural analogues reveals the unique nature of magnetic interactions in this Mn-based system. Compounds such as YCa$_3$(CrO)$_3$(BO$_3$)$_4$ (Cr$^{3+}$, $d^3$)~\cite{Wang7535} and YCa$_3$(VO)$_3$(BO$_3$)$_4$ (V$^{3+}$, $d^2$)~\cite{Silverstein044006,Miiller1315} exhibit dominant AFM interactions along their respective chains, with $\theta_{\rm CW} \simeq–120$~K and $–453$~K, respectively, and show no evidence of long-range magnetic ordering down to the lowest measured temperatures, indicating significantly stronger magnetic frustration in those systems. In contrast, similar to gaudefroyite, a natural mineral with the composition Ca$_4$Mn$_3$O$_3$(BO$_3$)$_3$CO$_3$, Mn-based YCa$_3$(MnO)$_3$(BO$_3$)$_4$ exhibits a remarkable coexistence of FM chains and frustrated AFM 2D kagome layers, making it a promising candidate for exploring the interplay between structure, orbital physics, and low-dimensional magnetism~\cite{Li21149}.

\subsection{Critical Analysis of Magnetization}
\begin{figure*}
\includegraphics[width=0.85\textwidth]{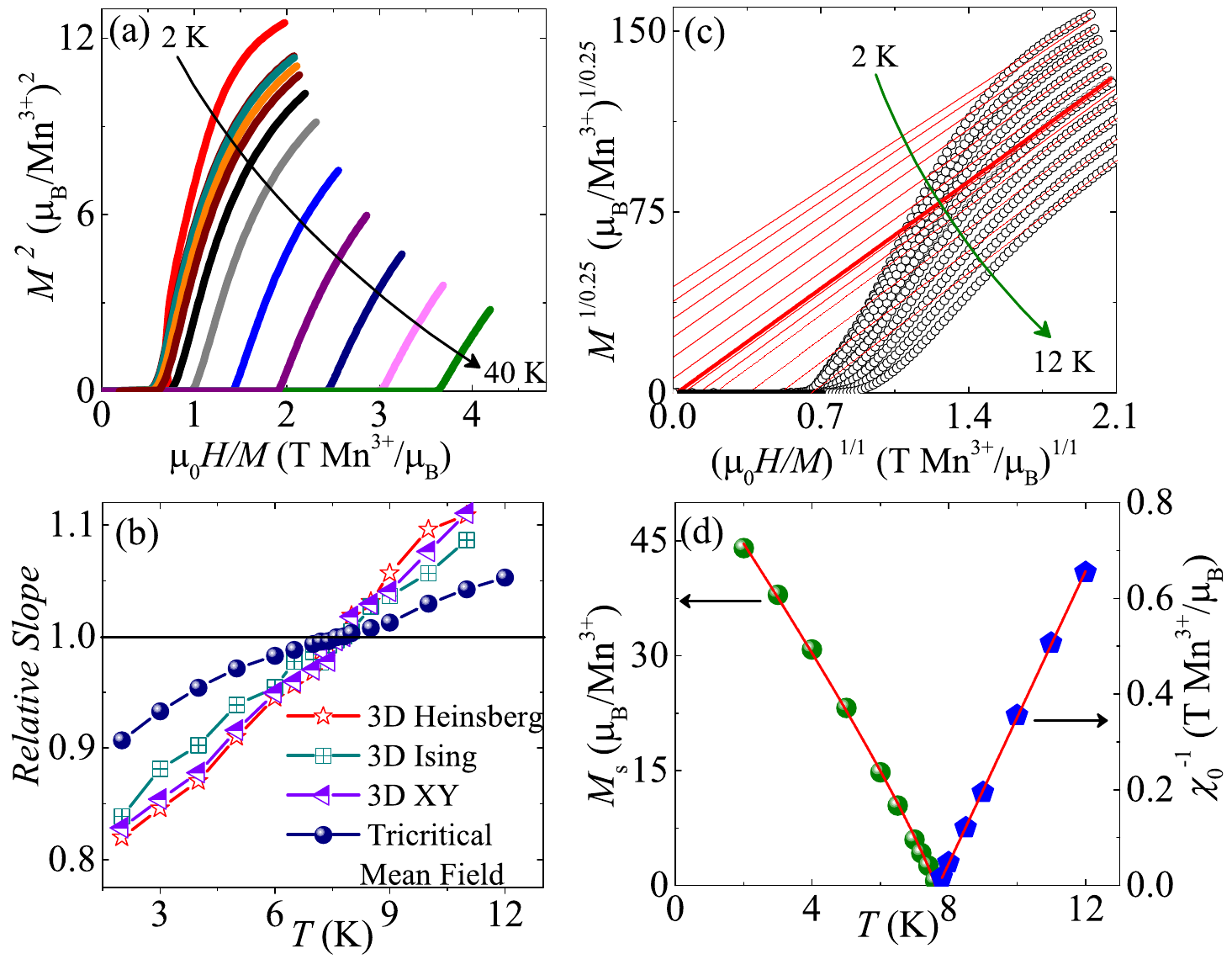}
\caption{\label{Fig9}(a) The Arrot plots at different temperatures near $T^*$. (b) Relative slope ($RS$) as a function of $T$ for different models mentioned in Table~\ref{CA_table}. (c) MAPs using the TMFM. The solid lines are the linear fits to data in high field regime. (d) Variation of spontaneous magnetization ($M_{\rm s}$) and zero field inverse susceptibility ($\chi_{0}^{-1}$) as function of $T$ in left and right axes with solid lines as a fit using the Eq.~\eqref{CA_Ms} and~\eqref{CA_chi0}, respectively.}
\end{figure*}
According to the mean field model, magnetic isotherms ($M$ vs $H$) recorded close to the transition temperature should follow~\cite{Arrott1394}, 
\begin{equation}
\label{MF}
H/M = a + bM^{2},
\end{equation}
where $a$ and $b$ are constants. The plot of $M^{2}$ vs $H/M$ at different temperatures are generally known as the Arrott plots. They are widely used in ferromagnets for a precise estimation of the ferromagnetic transition temperature and to understand the nature of magnetic transition~\cite{Arrott1394,Stanley1987}. The Arrott plots for YCMBO are shown in Fig.~\ref{Fig9}(a). According to the Banerjee criteria, one can determine the nature of a magnetic phase transition by analyzing the slope of the Arrott plots~\cite{Banerjee16}. Positive slope indicates a second-order phase transition (SOPT), while a negative slope implies a first-order phase transition (FOPT). The observed positive slope for YCMBO over the entire temperature range [Fig.~\ref{Fig9}(a)] signifies SOPT. However, Arrot plots don't follow Eq.~\eqref{MF}, as they are not linear, suggesting that the transition is non-mean field type.

In order to determine correct universality class of the phase transition, we adopted modified Arrott plots (MAPs), which are based on the Arrott-Noakes equation~\cite{Arrott786},
\begin{equation}
\label{MAP}
(H/M)^{1/\gamma} = A\epsilon + BM^{1/\beta}.
\end{equation}
Here, $A$ and $B$ are constants, $\epsilon= \frac{T-T{\rm ^*}}{T{\rm^*}}$ represents the reduced temperature, and $\gamma$ and $\beta$ are critical exponents. The critical exponents derived from MAPs should generally fall under one of the universality classes listed in Table~\ref{CA_table}. To select the suitable model for YCMBO, we used the Relative Slope ($RS$) method given in Ref.~\cite{Mazumdar093902}. Relative slope is defined as, $RS$ = $S(T)/S(T^*)$. Here, $S(T)$ and $S(T^*)$ are the slopes of the linearly fitted lines in the high-field region of the MAPs at different temperatures and at the transition temperature, respectively. The selection of the most accurate model for the given sample is based on how closely the $RS$ value approaches unity both above and below $T^*$ for a particular model~\cite{Mazumdar093902}. Figure~\ref{Fig9}(b) shows $RS$ calculated considering different models for YCMBO. It is evident that the Tricritical Mean Field Model (TMFM) shows the least deviation from unity in both the regions, below and above $T^*$~K.  

Generally, MAPs plotted by an accurate model should be a set of parallel straight lines, and the line corresponding to the transition temperature will pass through origin~\cite{Kolay224405,Singh6981}. Figure~\ref{Fig9}(c) shows MAPs using the TMFM critical exponents, $\beta = 0.25$ and $\gamma = 1$. They are almost straight lines in the high-field region. A linear fit is done in this range and extrapolated down to zero-field. It results in a set of parallel lines, and the line corresponding to $T_{\rm TMFM}=7.8$~K passes through origin [bold red line in Fig.~\ref{Fig9}(c)]. 

Generally, multiple iterations of MAPs are necessary to obtain reliable values of the critical exponents. For the next iteration, the values of $\beta$ and $\gamma$ are determined by the power law dependence of the critical exponent on the reduced temperature scale near the transition temperature as~\cite{Islam134433,Singh6981},
\begin{equation}
\label{CA_Ms}
    M_{S} (T) = M_{0}(-\epsilon)^{\beta}~\text{for}~\epsilon < 0, 
\end{equation}
~~~~~~~~~~~~~~~~~~~~~~~~~~~~~~and
\begin{equation}
\label{CA_chi0}
    \chi_{0}^{-1} (T) = \Gamma(\epsilon)^{\gamma}~\text{for}~\epsilon > 0.
\end{equation}
Here, $M_{\rm S}$ is the spontaneous magnetization and $\chi_0$ is the zero field susceptibility. $M_{\rm S}$ and $\chi_0^{-1}$ are determined using the $y$-intecepts and $x$-intercepts of the extrapolated linear fits to the MAPs above and below the transition temperature, respectively. $M_{\rm S}$ and $\chi_{0}^{-1}$ obtained from Fig.~\ref{Fig9}(c) are plotted in Fig.~\ref{Fig9}(d). The solid lines in Fig.~\ref{Fig9}(d) are fits to $M_{\rm S}$ and $\chi_{0}^{-1}$ using Eqs.~\eqref{CA_Ms} and \eqref{CA_chi0}, respectively. From these fits we got new values of $\beta^{'} = 0.87$ and $\gamma^{'} = 1$. By using these values, we have constructed new MAPs and performed a linear fit in the high-field region, and extrapolated the fits (not shown here). For $T > T_{\rm TMFM}$, the $x$-intercepts are coming negative implying negative values of $\chi_{0}^{-1}$. These unphysical values suggest the next interaction is not needed and the values obtained from first iteration, $\beta=0.25$ and $\gamma=1.00$ are the reliable values for YCMBO. Therefore, the accurate transition temperature of YCMBO according to this model is $T_{\rm TMFM}\sim 7.8$~K. This value is close to the one obtained from magnetization and heat capacity data. 

Further, critical exponent $\delta$ is calculated by using Widom relation given by~\cite{Widom3898},
\begin{equation}
\delta = 1 + \frac{\gamma}{\beta}.
\end{equation} 
Using the values of $\beta=0.25$ and $\gamma=1.00$, we obtained $\delta=5$. These values of critical exponents for YCMBO align with the TCFM model when compared to all the universality classes given in Table~\ref{CA_table}. Thus, we can conclude that YCMBO undergoes a magnetic transition at $T^*=T_{\rm TMFM}\sim 7.8$~K and belongs to the TCMF universality class. 

\begin{table*}
\caption{List of the critical exponents $\beta$, $\gamma$, $\delta$, and $T^{*}$ of YCMBO calculated from MAPs and theoretical values for different universality classes taken from Ref.~\cite{Islam134433}.}
\label{CA_table}
\begin{ruledtabular}
\begin{tabular}{c c c c c c c c c}
 Parameters & MAPs &  Tri Critical & Mean field & 3D Heisenberg & 3D XY & 3D Ising \\
  &(YCMBO)& Mean field & model & model & model&model\\
 \hline
$\beta$&0.25&0.25& 0.5&0.365&0.345&0.325\\
$\gamma$&1&1& 1&1.386&1.316&1.241\\
$\delta$&5 &5 & 3&4.8&4.8&4.82\\
$T{\rm^* }$~(K)&7.8&-&-&-&-&-&\\
\end{tabular}
\end{ruledtabular}
\end{table*}

\subsection{Magnetocaloric Effect}
\begin{figure*}
\includegraphics[width=\textwidth]{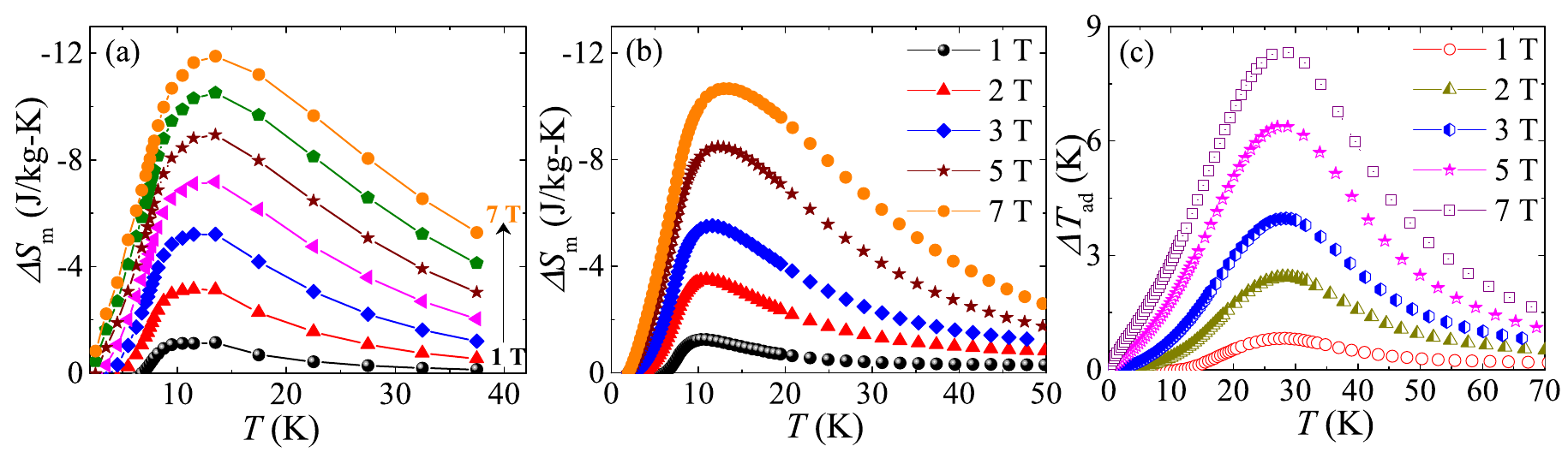}
\caption{\label{Fig10}(a) Isothermal magnetic entropy change ($\Delta S_{\rm m}$) as a function of $T$ calculated at various applied fields using Eq.~\eqref{Sm_1}. (b) Variation of $\Delta S_{\rm m}$ with $T$ calculated using field-dependent heat capacity data and Eq.~\eqref{Sm_2}. (c) Adiabatic temperature change ($\Delta T_{\rm ad}$) as a function of $T$ calculated at various applied fields using Eq.~\eqref{Tad_1}.}
\end{figure*}

\begin{figure*}
\includegraphics[width=\textwidth]{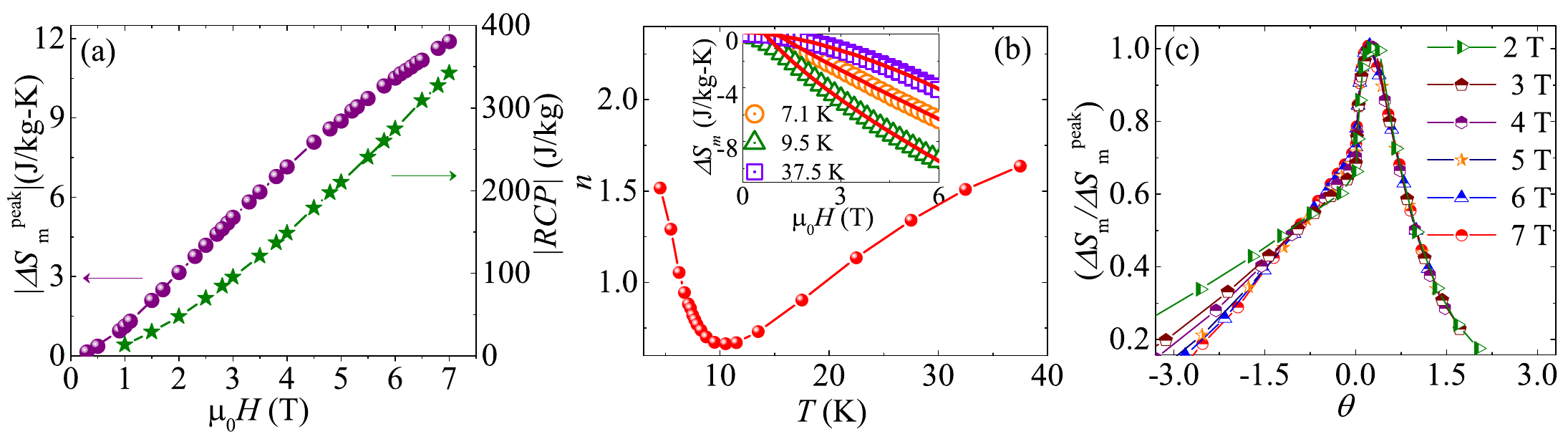}
\caption{\label{Fig11}(a) Field evaluation of $\Delta S_{\rm m}^{\rm peak}$ (left-axis) and $RCP$ (right-axis).
(b) The exponent $n$ as
a function of temperature, obtained from the ﬁtting of power law to $\Delta S_{\rm m}$ vs $H$ isotherms. A few representative fits are shown in the inset for temperatures near and well above $T_{\rm C}$. (c) Normalized magnetic entropy ($\Delta S_{\rm m}$/$\Delta S_{\rm m}^{\rm peak}$) are plotted as a function of the rescaled temperature ($\theta$) [Eq.~\eqref{universal_curve}] in different fields.}
\end{figure*} 

\begin{table}
\caption{MCE parameters of YCMBO are compared with previously studied materials having $T_{\rm C}$ (or $T_{\rm N}$) in the same temperature range. $\Delta H$ stands for the field change during the cooling cycle.}
\begin{tabular}{c c c c c c c}
\hline\hline
 Compounds &$T_{\rm C}$ or $T_{\rm N}$ & $\Delta{H}$ & 
  $|\Delta S_{\rm m}^{\rm peak}|$&  $RCP$ &  Ref.  \\
     & (K)&(T) & (J/kg-K) & (J/kg) \\
\hline
Tm$_{2}$FeCrO$_{6}$&  10.5 K &5 &  4.78  & 123.5 &  \cite{DONG26632}\\       
Er$_{2}$FeCr$_{6}$& 11.7/5.7 K&5 &11.95 & 215.8 &  \cite{DONG26632}\\
Er$_{2}$Ni$_{2}$Ga & 7.1 K &7 & 12.93 & 265 & \cite{Dan1}\\
HoMnO$_{3}$ &  5 K& 7 &13.1 & 320 &  \cite{Midya142514}\\
Tb$_3$RuO$_7$ &  12 K &5 & 7.78& 128.8 &  \cite{Ningzhou36968}\\
Gd$_{3}$RuO$_{7}$ &  11 K & 5 &8.12 & 124 &  \cite{Ningzhou36968}\\
YCMBO &  7.8 K & 7 &12 & 349 &This work\\ 
  \hline\hline
\label{Table_MCE}
\end{tabular}
\end{table}
 The magnetocaloric effect (MCE) is a property of magnetic materials where the temperature changes when subjected to the changing applied magnetic field. This is a useful property to achieve temperatures down to the milli-kelvin range. The protocol is to first apply a magnetic field to the sample isothermally and then remove the field slowly and adiabatically. The temperature of the sample is lowered in the second stage, and hence, this method is also widely known as adiabatic demagnetization refrigeration. In order to identify suitable materials for refrigeration, MCE is quantified by using two major parameters, change in isothermal entropy in the first stage ($\Delta S_{\rm m}$) and change in adiabatic temperature ($\Delta T_{\rm ad}$) in the second stage with respect to the change in applied field ($\Delta H$).
 
 We calculated $\Delta S_{\rm m}$ separately by using magnetization isotherms and heat capacity data. In the first method, magnetization isotherms ($M$ vs $H$) were recorded in close temperature steps in the vicinity of $T^*$. Then, Maxwell thermodynamic relation, $(\frac{\partial{S}}{\partial{H}})_{T}$ = $(\frac{\partial {M}}{\partial {T}})_{H}$ is integrated to get
 \begin{equation}
\label{Sm_1}
\Delta S_{m}(H,T) = \int_{H_{i}}^{H_{f}}\frac {dM}{dT} \,dH,
\end{equation}
where, $dM$ is the difference between the magnetization of two curves apart by temperature $dT$. The obtained $\Delta S_{\rm m}$ as a function of temperature is shown in Fig.~\ref{Fig10}(a) for different values of $\Delta H = H_{f} - H_{i}$ from 1~T to 7~T. All the curves increase with temperature and show a broad maxima around 13~K. The maximum peak value of  $\Delta S_{\rm m}\simeq 12$~J/kg-K is observed for a field change of $\Delta H$= 7 T.

Further, to cross-check the values of $\Delta S_{\rm m}$, we have also done the calculations using field-dependent heat capacity data. First, the total entropy at a given field $H$ is obtained by simply integrating $(C_{\rm p})_H$ as, $S(T)_{H} = \int_{T_{i}}^{T_{f}}\frac {(C_{\rm p})_H}{T}dT$. Next, $\Delta S_{\rm m}$ is obtained by subtracting the zero-field entropy from the entropy at higher fields as,  
\begin{equation}
\label{Sm_2}
\Delta S_{\rm m}(T) = S(T)_{H} - S(T)_{0}.
\end{equation}
Figure~\ref{Fig10}(b) presents the obtained $\Delta S_{\rm m}$ as a function of temperature in different fields. The overall shape of the curve and maxima position are the same as obtained by the first method, with only a slight reduction in the peak values. The value of $\Delta S_{\rm m}$ calculated using the heat capacity data is regarded to be more accurate than that obtained using the magnetization data~\cite{Magar054076}. The maximum value of $\Delta S_{\rm m}$ obtained at 7~T is $\sim 11$~J/kg-K. Likewise, $\Delta T_{\rm ad}$ is calculated by taking the difference in temperature corresponding to the same entropy (adiabatic change) at two different applied fields as~\cite{Magar054076}, 
\begin{equation}
\label{Tad_1}
\Delta T_{\rm ad} = T(S)_{H_{f}}-T(S)_{H_{i}}.
\end{equation}
Figure~\ref{Fig10}(c) depicts the variation of $\Delta T_{\rm ad}$ as a function of temperature. The maximum value of $\Delta T_{\rm ad}$ is found to be around 8.4~K for $\Delta H = 7$~T. 

As we got large values of MCE parameters, we also calculated one more figure of merit to determine the potential of YCMBO for magnetic refrigeration application, namely relative cooling power ($RCP$). $RCP$ quantifies the amount of heat transfer between the cold and hot reservoirs in the refrigeration cycle, which can be expressed mathematically as
\begin{equation}
RCP = \int_{T_{\rm cold}}^{T_{\rm hot}}\Delta S_{m}(T,H) \,dT.
\end{equation}
The approximated formula for $RCP$ is given by
\begin{equation}
|RCP|_{\rm simeq} = \Delta S_{\rm m}^{\rm peak} \times \delta T_{\rm FWHM},
\label{RCP}
\end{equation}
where $\Delta S_{\rm m}^{\rm peak}$ is the peak value and $\delta T_{\rm FWHM}$ is full width half maxima of the $\Delta S_{\rm m}$ vs $T$ curves in Fig.~\ref{Fig11}(a). $\Delta S_{\rm m}^{\rm peak}$ and the calculated $RCP$ as a function of field are shown in Fig.~\ref{Fig11}(a). The maximum value of $RCP$ is obtained to be $\sim 349$~J/kg for 7~T.

For practical application of the material in magnetic cooling, large values of $\Delta S_{\rm m}$, $\Delta T_{\rm ad}$, and $RCP$ are not only the necessary criteria. The nature of magnetic transition also plays a decisive role. For instance, materials with FOPT might have large peak heights of the $\Delta S_{\rm m}$ curves, but their width is narrow. Also, these transitions are non-reversible and come with thermal hysteresis~\cite{Franco305}. Due to these limitations, FOPT materials have less working range and efficiency in the cyclic processes of MCE. On the other hand, materials undergoing SOPT with a reversible nature and negligible hysteresis are more desirable. In order to examine the nature of the phase transition, we followed the method given in Ref.~\cite{Singh6981}. The plots of $\Delta S_{\rm m}$ vs $H$ for different temperatures near the transition are fitted by a power law, $\Delta S_{\rm m}\propto H^{n}$ as shown in the inset of Fig.~\ref{Fig11}(b). The obtained $n$ values as a function of temperature are plotted in Fig.~\ref{Fig11}(b). The value of $n$ remains below 2, implying SOPT~\cite{Mohanty134401,Law2680}.

Franco et. al.~\cite{Franco222512} proposed another method to determine the nature of the phase transition from the construction of universal scaling curve of $\Delta S_{\rm m}$. Universal scaling curve is constructed by plotting $\Delta S_{\rm m}$/$\Delta S_{\rm m}^{\rm peak}$ (normalized $\Delta S_{\rm m}$) vs $\theta$. Here, $\theta$ is the rescaled temperature axis. The rescaling is done as,
\begin{equation}
\theta= \begin{cases}-(T-T^*)/(T_{\rm r1}-T^*),~T\le T^* \\\\-(T-T^*)/(T_{\rm r2}-T^*),~T\ge T^*,
\end{cases}
\label{universal_curve}
\end{equation}
where, $T_{r1}$ and $T_{r2}$ are two temperature extrema corresponding to half of $\Delta S_{\rm m}^{\rm peak}$. If all the scaled curves collapse to a single curve (also known as the master or universal curve), that implies SOPT~\cite{Bonilla09E131,Dong116101}. Figure~\ref{Fig11}(c) presents the universal curve construction for YCMBO. All curves in different fields are nicely overlapping with each other, further confirming SOPT in YCMBO.

We have compared the MCE parameters of YCMBO with already reported MCE materials having the magnetic transition near to $T^*$ in Table~\ref{Table_MCE}. The MCE parameters of YCMBO are comparable to those of the other potential candidates. Additionally, it has a considerably large value of $RCP$, among the listed compounds. It is worth noting that YCMBO is the only rare earth-free compound in Table~\ref{Table_MCE}. YCMBO has three $S=2$ magnetic Mn$^{3+}$ ions having the potential to accommodate large zero-field entropy of about $S_{\rm m} =40$~J/mol~K. Further, due to major FM interaction between the ions, we can observe considerable entropy change with a small change in magnetic field. The competing FM-AFM interaction near the transition causes the entropy to distribute over a large temperature range as reflected in broad $\Delta S_{\rm m}$ vs $T$ curves. Therefore, YCMBO gives rise to a substantial $RCP$ value. With large MCE parameters, SOPT, and negligible hysteresis YCMBO is found to be yet another potential material for magnetic refrigeration.

\section{Summary}
In summary, we revisited the magnetic properties and studied the MCE character of YCMBO in detail. The kagome lattice compound undergoes a FM ordering at $T^* \simeq 7.8$~K with a spin canting due to competing FM and AFM interactions. A power law fit to $C_{\rm mag}/T$ below $T^*$ gives an exponent $n \simeq 0.5$, indicative of a FM transition. Despite coexisting FM and AFM interactions, the frustration parameter ($f \simeq 4$) is found to be considerably large, a robust signature of strong frustration in the system. Band structure calculations confirm dominant FM interaction along the chain and weak AFM interaction in the kagome plane perpendicular to the chain direction. It exhibits a large MCE ($\Delta S_{\rm m} \simeq 12$~J/kg-K, $\Delta T_{\rm ad} \simeq 8.4$~K, and $RCP \simeq 349$~J/kg for $\Delta H = 7$~T.) comparable to other reported MCE materials having the transition temperature at around $T^*$. The critical analysis of magnetization via modified Arrott plots and MCE parameters establish SOPT at $T^*$ and a tricritical mean field type FM with critical exponents $\beta = 0.25$, $\gamma = 1$, and $\delta = 5$. Being a rare-earth free compound, along with large MCE parameters, SOPT, and absence of magnetic hysteresis, YCMBO serves as an excellent material for cooling purpose using ADR technique.

\section{acknowledgments}
For financial support, we would like to acknowledge SERB, India bearing sanction Grant No.~CRG/2022/000997 and DST-FIST with Grant No.~SR/FST/PS-II/2018/54(C).


%

\end{document}